# Numerical Simulation of Quantized Current Generated by a Quantum Dot Pump

Ye-Hwan Ahn[1, 2] and Yunchul Chung[3, a]

[1)] Korea Research Institute of Standards and Science, Daejeon 34113, Republic of Korea

[2)] Department of Physics, Korea University, Seoul 02841, Republic of Korea

[3)] Department of Physics, Pusan National University, Busan 609-735, Republic of Korea

[a)] ycchung@pusan.ac.kr

(Dated: 8 September 2017)

The quantized current generated by a quantum dot pump is calculated numerically. The numerical simulation is done by dividing the time varying potential into many static potentials with a short time interval and calculating the electron capture and pumping rate with the time independent Schrödinger equation. The simulation results show good agreement with reported experimental results qualitatively. The calculated 2D pump current map and the plateau width dependence on the modulation gate voltage show good agreement with the experimental results. From the simulation results, it is explained how the back-tunneling process affects the accuracy of the current plateaus quantitatively. Also, the energy distribution of the pumped electron is calculated, which can be measured experimentally. Finally, it is found that the pump current accuracy can be enhanced by increasing the entrance gate width, which is important to realize the quantum current standard.

## I. INTRODUCTION

The single electron pump based on a quantum dot (QD) is attracting attention as a promising device[1–7] to realize the future quantum current standard.[8–12] Also, spin pumping in graphene devices[13–18] and adiabatic pumping[19–21] were studied recently. It has been reported that the quantized current pumping can be achieved by applying periodic voltage modulation on a quantum dot gate.[3–5] The pumping mechanism has been studied theoretically with an assumption that time varying electron states can be considered as stationary states by dividing the modulation period into short enough time intervals.[4] This approach allows consideration of the problem by solving the time-independent Schrödinger equation consecutively instead of solving the Floquet scattering problem.[22]

In order to use the QD pump as a current standard, the current accuracy should be as precise as the $10^{-8}$ level.[23] The accuracy of the pumped current depends on many experimental parameters such as the modulation frequency, energy level spacing inside a QD, and the detail geometrical shape of the QD gates and so on. To understand the influence of these parameters on the current accuracy, it is useful to calculate the so-called 2D pump current map as a function of entrance and exit gate voltages for various experimental parameters mentioned above. However, such detail analysis based on the 2D pump current map has not been done yet.

In this work, we have calculated the 2D pump current map for various experimental parameters by numerically solving the stationary state problem mentioned above. Our calculation is based on the non-interacting model. We assume that the tunneling rate of an electron is not influenced by the correlation between electrons in the dot. Also, we assume that the QD potential is only determined by the gate voltages of the QD and not affected by the potential of electrons in the dot. The calculated 2D pump current map shows good agreement with the experimental result qualitatively. The calculated results for various entrance gate widths indicate that the accuracy of the pumped current can be enhanced by increasing the entrance gate width. Even though our model is based on a simple Schrödinger equation solver, it helps us to understand most of the important pumping mechanisms intuitively.

## II. NUMERICAL SIMULATION

Figure 1(a) shows the schematic potential of the quantum dot pump, fabricated on a 1D wire, used for our calculation. We adopted a simple quantum model of non-interacting electrons confined in a one-dimensional wire, used by Kaestner and Kashcheyevs.[4]

The device model, used for calculation, is based on the real device geometry used in experiments, which employs two metallic gates on a conducting channel etched in a two-dimensional electron gas (2-DEG).[1, 5] When a time varying voltage ($V_{AC}$) is applied to the fixed entrance gate voltage ($V_{ENT}$), the QD potential changes dynamically in time as shown in Fig. 1(a). The pumping cycle can be categorized into four phases: (i) loading, (ii) back-tunneling, (iii) trapping, and (iv) ejection.[7] A proper choice of the modulation voltage amplitude will allow the QD (formed between entrance and exit gates) to capture electrons below the left Fermi sea (potential profile i) and transfer to the right side of the QD (potential profile iv).

The pumped current is calculated by using the rate equation for the ground state energy level occupation probability $P(t)$[4]

$$\hbar \frac{dP}{dt} = (\Gamma^L + \Gamma^R)[f_F(E_1) - P] \quad (1)$$

Here, $E_1$ is the QD ground state energy at time $t$, $f_F$ is the Fermi function, and $\Gamma^L$ and $\Gamma^R$ are the corresponding electron tunneling rates to the left and right sides of the QD, respectively. We limit our calculation to absolute zero temperature. During the loading stage (when the QD confinement energy is below the Fermi sea, i.e., $E_1 < E_F$), the probability of the electron loading into the dot during the time interval $dt$ can be approximated to $\Gamma^L(1-P)dt$, when $\Gamma^L \gg \Gamma^R$. The condition usually holds because the exit barrier is much thicker and higher than the entrance barrier, which can be easily checked from the calculated tunneling rates (region i) in Fig. 1(b). As soon as the QD confinement energy is lifted above the Fermi energy ($E_1 > E_F$), the confined electron in the QD starts to leak backwards with the rate of $\Gamma^L P dt$. This is one of the most dominant processes that deteriorate the quantized current accuracy (region ii). The backward tunneling decreases rapidly as the left barrier is further lifted due to the exponential decrease in the tunneling rate (region iii). As the entrance barrier is lifted above the exit barrier, the electron starts to eject to the right side of the QD with the rate of $\Gamma^R P dt$. By integrating the loading, back-tunneling, and ejection of the electron throughout the pump cycle, the pump current can be calculated.

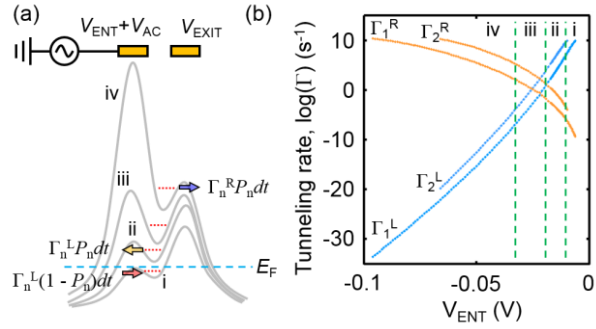

FIG. 1. (a) Schematic diagram showing potential changes in the QD pump during the pumping cycle. The thick blue lines on top represent the entrance and exit gates on the 1D wire surface. Due to the periodic modulation of entrance gate potential (by A.C. voltage $V_{AC}$ added on the fixed entrance gate voltage $V_{ENT}$), the QD potential, especially the entrance gate potential, changes dynamically in time. (b) The calculated tunneling rates of the confinement states in the QD as a function of entrance gate voltages. $\Gamma_n^L$, $\Gamma_n^R$ are the tunneling rates to the left and right sides of the QD for the $n$-th QD confinement energy states, respectively.

The rate equation Eq. (1) is valid when a single electron is pumped via the ground state of the QD, and the occupation of the considering energy state can be determined only by the tunneling through the dot, not by the transition between different energy states. In a pump device, the transition between different energy states is quite negligible because the electrons in the higher energy states always get emptied by tunneling before the lower energy states are available for significant relaxation due to the large difference in the tunneling rates between higher and lower energy states [see Fig. 1(b), $\Gamma_2$ is typically at least more than 1000 times bigger than $\Gamma_1$, at the same entrance gate voltage.]. Therefore, the occupation change due to the transition is quite well suppressed. Hence, we used the same rate equation to consider the electron pumping through higher energy states just to calculate the qualitative behavior of the higher current plateaus ($I/ef > 1$). This extension will not give quantitative results like the first plateau but will allow us to study some fundamental aspects of the QD pump, with reasonable computation time. It took 12 h to calculate a 2D pump current map by using a desktop computer with the voltage resolution shown in this paper.[24]

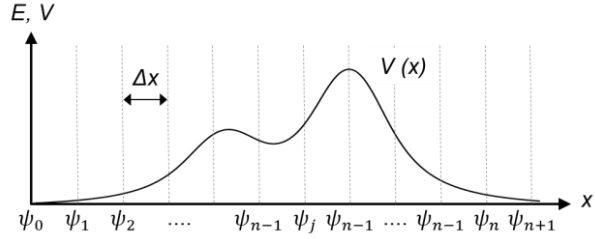

FIG. 2. Schematic diagram showing the discretization of the 1D potential profile of the pump device and the wave function.

The calculation was carried out by the following steps: First, the electron tunneling path along the 1D conducting channel was represented as $\Delta x$=5 nm sized 1D mesh-points to allow numerical calculation, as it is shown in Fig. 2. The QD potential profile $V(x)$, which is static at a given time $t$, was calculated using the analytical solution developed by Davies et al.[25] Second, the tunneling probability as a function of energy was calculated by solving the time-independent effective-mass Schrödinger equation numerically[26] using the finite-difference method (FDM). The Schrödinger equation is discretized at a given lattice point j as follows:

$$-s\psi_{j+1} + 2s(V_j - E)\psi_j - s\psi_{j-1} = 0 \qquad (2)$$

Here, $s$ is $\frac{\hbar}{2m^*\Delta x^2}$, where $m^*$ is the effective mass of the electron in the GaAs conduction band. The quantum transmitting boundary method[27] is used to set the boundary conditions for the left and the right end of the mesh-points. The full calculation details, including the discretization of the Schrödinger equation, application of the boundary condition, as well as numerically techniques, can be found in the paper reported by Fernando and Frensley.[26]

Because the tunneling rate cannot be obtained directly by solving the Schrödinger equation with FDM, it was obtained by fitting the energy dependency of the tunneling probability near the resonance with the Breit-Wigner formula [28] of Eq. (3). From the fitting, two values of Γ can be obtained. However, these values can be assigned to

$\Gamma^L$ and $\Gamma^R$ only after evaluating the transmission probabilities (from the center of the dot) to the left and right sides of the QD, which can be calculated by solving the Schrödinger equation. If the transmission probability is higher to the left side, the higher tunneling rate is assigned to $\Gamma^L$, while the lower tunneling rate is assigned to $\Gamma^R$ and vice versa (because the tunneling rate is proportional to the tunneling probability)

$$T(E) = \frac{\Gamma^L \Gamma^R}{(E-E_1)^2 + \left(\frac{\Gamma^L+\Gamma^R}{2}\right)^2} \tag{3}$$

Repeating the above steps for different entrance gate voltages, the tunneling rates are calculated as shown in Fig. 1(b). The fitting is quite reliable in most cases except for the case with the extremely small energy resonance width $h(\Gamma^L + \Gamma^R)$ (usually when the resonance width becomes below $10^{-15}$ eV), which hardly contributes to the pumping current calculation. Once the tunneling rates are calculated for the entrance gate voltages of interest (from $V_{ENT} - V_{AC}$ to $V_{ENT} + V_{AC}$), the pump current is calculated by integrating the rate equation [Eq. (1)] for the whole pump cycle. In a real device, the energy difference between the ground and the first excited state is mostly determined by the QD charging energy (which is around a few meV) because the energy level spacing is usually a few times smaller than the charging energy. However, it is not possible to incorporate the charging energy of a QD into the Schrödinger equation directly. Hence, we adjusted the width of our potential well so that the energy level spacing is comparable to a real device with a typical charging energy (2–3 meV).

## III. RESULTS AND DISCUSSION

The calculated result clearly shows good qualitative agreements with the reported experimental results,[4, 29] as it is shown in Fig. 3(a). The clear current plateaus for $I/ef$=1 and 2 (marked as A and B in the figure, respectively) are shown. These plateaus appear when the number of electrons loaded in the QD is the same as the number of electrons ejected from the QD, while plateau C ($I/ef$=1) appears when two electrons are loaded in the QD and only one electron is ejected. These features are experimentally observed and well explained.[30] To check the reliability of our simulation result, the calculated current is compared [Fig. 3(b)] with the decay cascade model which gives an analytic expression for the quantized current of the QD pump.[28] The calculated current shows a reasonable fit to the formula proposed by the decay-cascade model. The first current plateau ($I/ef$=1) fits quite nicely to the model, while some deviation is observed for the second current plateau. This is probably because our model does not consider the relaxation and other details during the initialization process, while the decay-cascade model does. Also, it is because the ratio of $\Gamma_n/\Gamma_{n-1}$ is assumed to be constant in the decay-cascade model, while in our model, the ratio slightly changes as a function of the entrance gate voltage. This is due to the electrostatic effects of the gate voltage on QD potential, as it can be seen in Fig. 1(b). Nevertheless, we believe that the result is reasonable enough to study the qualitative behavior of the pumping mechanisms and so on.

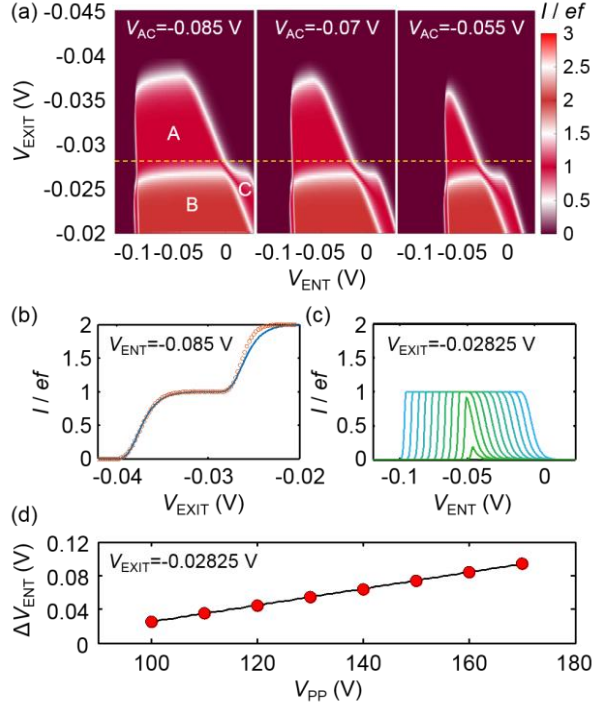

FIG. 3. (a) Calculated 2D pump current map as a function of exit and entrance gate voltages. For the simulation, a 70 nm deep 2DEG with a Fermi energy of 7 meV is used, and the width of the entrance and the exit gates was set to 150 nm with a gap of 150 nm between them. The entrance gate was modulated with 2 MHz sinusoidal A.C. voltage. The modulation voltage amplitudes are 0.085 V, 0.07 V, and 0.055 V from left to right. (b) The calculated quantized current as a function of an exit gate voltage, when the entrance gate voltage is set to -0.085 V. The result (red hollow circle) is fitted with a decay-cascade model (blue line).[6] (c) The calculated plateau width dependence on the modulation voltage. The current plateau widths are measured at -0.02825 V of exit gate voltage [dashed line in Fig. 1(a)]. The modulation amplitude is changed from 0.085 V (widest plateau) to 0.03 V (tiny peak around VENT = -0.04 V) in 0.005 V steps. (d) The dependence of the current plateau width to the modulation voltage (peak-to-peak voltage, $V_{PP}$).

The plateau width dependence on the modulation voltage is calculated as shown in Fig. 3(c). The plateau width (in entrance gate voltage) is calculated by changing the modulation voltage and estimating the first plateau width at a fixed exit gate voltage [dashed line in the Fig. 3(a)]. The plateau width linearly increases [see Fig. 3(d)] for higher modulation voltages, as it was reported by Kaestner et al.[5] Even though we used a simple model for the simulation, the results match quite nicely with the reported experimental results (quantitatively with the first plateau and qualitatively with the second plateau).

The energy distributions of the back-tunneling and the pumped electrons are calculated at the beginning and the end of the transition region between the first ($I/ef = 1$) and the second ($I/ef = 2$) plateau in Fig. 4. At the green point where the current starts to deviate from $I/ef = 2$, the electron in the second confinement states starts to leak backwards noticeably [peak C in Fig. 4(c)] as soon as the second confinement energy state is lifted above the Fermi energy ($E_2 > E_F$). The back-tunneling rate decreases exponentially as the entrance gate is further lifted [see $\Gamma_2^L$ in Fig. 1(a)], which explains the rapid decrease in the back tunneling electron. At this exit voltage, the tunneling rate is not high enough to totally empty the dot due to the thick entrance barrier. Hence, most of the electron is pumped to the right side of the QD (peak B).

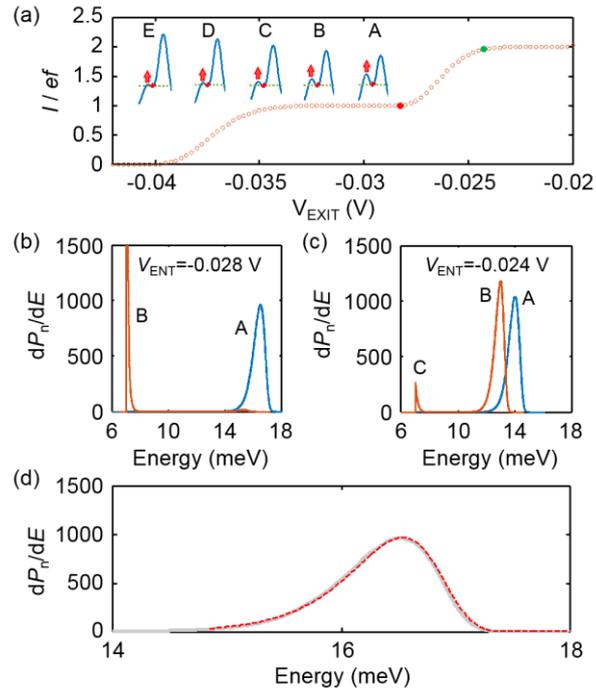

FIG. 4. The energy distributions of the back-tunneling and pumped electrons. (a) Two points where the energy distributions are calculated. The calculated QD potentials and the ground state energies (upper part) for $V_{EXIT}$ = -0.04 V, -0.037 V, -0.034 V, -0.031 V, and -0.028V (from left to right), when the ground state is lifted just above the Fermi energy, are shown. (b) and (c) energy distributions calculated at the red and green points, respectively. The blue and the red line represent the energy distributions of the tunneling electron from the first and the second confinement energy ($E_1$ and $E_2$) states, respectively. (d) Peak A fitted with Eq. (4) (orange dashed line).

The situation becomes quite the opposite at the onset of the transition between the first and the second plateau (hereafter the second transition), where the pumped current gets closer to $I/ef = 1$ [red point in the figure (a)]. At this point, the electron starts to fill the second energy state just below the Fermi energy (because the second confinement energy gets lower than the potential maximum of the entrance barrier just below the Fermi energy). Hence, when the state is lifted above the Fermi energy, the energy state is still very close to the top of the entrance barrier as it is shown in Fig. 5(a). Thus, the tunneling rate to the source is considerably high ($\Gamma_2^L \sim 3.1 \times 10^9$/s). Such a high tunneling rate will allow the electron to tunnel out from the QD rapidly (order of $10^{-9}$ s). This tunneling event is shown as sharp back-tunneling peak B in Fig. 4(b) (out of the scale). Hence, little electron is pumped even though the electron is loaded in the QD below the Fermi energy. The above results can be summarized that the onset of the transition to the *n*-th current plateau appears when the electron starts to fill the *n*-th energy state just below the Fermi energy during the pump cycle.

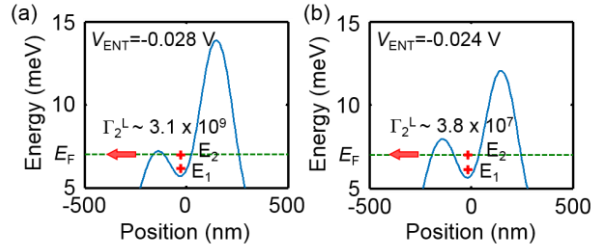

FIG. 5. (a) and (b) Calculated QD potentials and confinement energy states in the QD when the second confinement energy ($E_2$) states are aligned just above the Fermi energy at the red and the blue points, respectively.

Peak A in Fig. 4(b) and peaks A and B in Fig. 4(c) show the asymmetric energy distribution of the pumped electron. Fujiwara et al. observed asymmetric peaks in the transition regions when the pump current is differentiated with respect to the exit gate voltage. The asymmetric peaks are explained by the exponential dependence of the back-tunneling rate in energy.[3] The same principle holds for the energy distribution of the pumped electron. Since the tunneling rate to the right side of the QD is exponentially increasing during the pump cycle [see Fig. 1(b)], the tunneling rate can be written as $\Gamma(E) = \Gamma_1 e^{(E-E_1)/E_1}$, where $\Gamma(E_1) = \Gamma_1$. With the rate equation [Eq. (1)], the energy distribution of the pumped electron can be written as follows:

$$\frac{dP(E)}{dE} = c_1 \Gamma_1 \left[ e^{(\theta - E_1 \Gamma_1 e^\theta)} \right] \qquad (4)$$

Here, $\theta = (E - E_1)/E_1$ and $c_1$ is a constant. The numerically calculated energy distribution of the pumped electron (peak is fitted with the above equation in Fig. 4(d), which shows the perfect match. Such an asymmetric energy distribution of the pumped electron can be experimentally measured by using a similar method used by

Fletcher et al. under a high magnetic field.[31]

Figure 6(a) shows the current plateaus calculated for various entrance gate widths. The plateaus become wider and more horizontal as the entrance gate width increases. Also, the deviation of the plateau current from the expected current ($\Delta I$=1 - $I/ef$) reduces as the entrance gate width increases [Fig. 6(b)]. At point A in Fig. 4(a), the ground state is located far below the entrance potential maximum. Here, the back-tunneling is negligible due to the thick entrance barrier for the ground state energy. As the exit gate voltage becomes more negative (A→E), the depth of the QD becomes shallower and the ground state locates closer to the entrance potential maximum, which makes the entrance barrier thinner. Hence, the deviation DI increases as the exit gate voltage becomes more negative. The rate of the deviation increase can be slowed by introducing a thicker entrance barrier, hence making the pumped current more accurate.

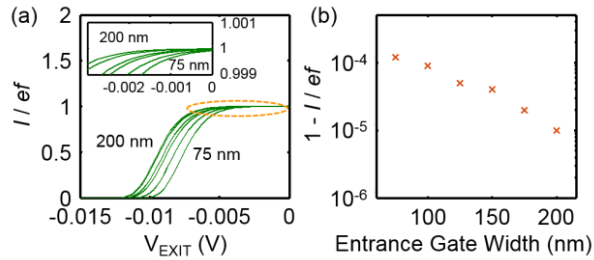

FIG. 6. (a) The first current plateau calculated for various entrance gate widths [200 nm (left) to 75 nm (right) in 25 nm steps]. Each plot was subtracted with different offsets for comparison. The inset shows that the slopes of the current plateaus vary with the entrance gate widths. (b) The deviation of the current (1 – $I/ef$) at the minimum point of $dI/dV_{EXIT}$ for various entrance gate widths.

This scenario assumes that the current accuracy is determined only by the back-tunneling, according to our simple model. However, the current accuracy can also be influenced by the electron loading process. During the electron loading stage (when the QD is below the Fermi energy), electrons can be partially loaded or over loaded in the QD if the electron loading process is not adiabatic (especially at a high modulation frequency). Our simple model cannot consider such non-adiabatic processes. Until now, it is unclear which process (either back-tunneling or the loading accuracy) is more dominant in determining the current accuracy. Hence, it would be interesting to test our proposal experimentally to determine the contribution of back-tunneling to the current accuracy.

## IV. CONCLUSION

The quantized current generated by a quantum dot pump was calculated numerically. The results show good qualitative agreement with reported experimental results. Even though the model is based on a simple Schrödinger equation solver, it provides intuitive understanding about the pumping mechanism of the QD pump. The energy distribution of the pumped electron and the pump current accuracy dependence on the entrance gate width are calculated, which can be tested experimentally. Our numerical model can be enhanced by adopting more realistic QD potential, which can be calculated by solving Poisson-Schrödinger equations self-consistently. This will enable the quantitative simulation of the QD pump current.

## ACKNOWLEDGMENTS

This work was supported by a 2-year Research Grant of Pusan National University. We thank N. Kim for helpful discussion.